\documentstyle[aps,prl,multicol,amssymb]{revtex}
\title{The non-local content of quantum operations}

\author{Daniel Collins,$^{1,2}$ Noah Linden$^3$ and
Sandu Popescu$^{1,2}$}

\address{ $^1$H.H. Wills Physics Laboratory,
University of Bristol, Tyndall Avenue, Bristol BS8 1TL, UK
\\ $^2$BRIMS, Hewlett-Packard
Laboratories, Stoke Gifford, Bristol BS12 6QZ, UK\\$^3$Department
of Mathematics, University of Bristol, University Walk, Bristol
BS8 1TW, UK}

\date{24 May 2000}
\begin{document}
%\draft
\maketitle
\begin{abstract}
We show that quantum operations on multi-particle systems have a
non-local content; this mirrors the non-local content of quantum
states.  We introduce a general framework for discussing the
non-local content of quantum operations, and give a number of
examples.  Quantitative relations between quantum actions and the
entanglement and classical communication resources needed to
implement these actions are also described.  We also show how
entanglement can catalyse classical communication from a quantum
action.

\end{abstract}


%\begin{multicols}{2}
\newcommand{\tr}{\mbox{Tr} }
\newcommand{\ket}[1]{\left | #1 \right \rangle}
\newcommand{\bra}[1]{\left \langle #1 \right |}
\newcommand{\amp}[2]{\left \langle #1 \left | #2 \right. \right \rangle}
\newcommand{\proj}[1]{\ket{#1} \! \bra{#1}}
\newcommand{\ave}[1]{\left \langle #1 \right \rangle}
\newcommand{\superop}{{\cal E}}
\newcommand{\unity}{\mbox{\bf I}}
\newcommand{\hilbert}{{\cal H}}
\newcommand{\relent}[2]{S \left ( #1 || #2 \right )}
\newcommand{\banner}[1]{\bigskip \noindent {\bf #1} \medskip}
%%%  Mathematical abbreviations  %%%
\newcommand{\I}{{\mathbf I}}
\newcommand{\R}{{\mathbf R}}
\renewcommand{\S}{{\mathbf S}}
\newcommand{\up}{\uparrow}
\newcommand{\down}{\downarrow}

\begin{multicols}{2}

\section{Introduction}

In the past,  most of the  research on quantum non-locality has
been devoted to the issue of non-locality of {\it quantum states}.
However we feel that an equally important issue is that of
non-locality of {\it quantum evolutions}. That is, in parallel
with the understanding of  non-locality of quantum kinematics one
should also develop an understanding of the non-locality of
quantum dynamics.

Let us start with a simple example. Consider two qubits situated
far from each other, one held by Alice and the other one by Bob.
Suppose they would like to implement a two qubit quantum evolution
described by the unitary operator $U$. (We wish   to be able to
apply $U$ on {\it any} initial state of the two qubits). With
exception of the case when $U$ is a product of two local unitary
operators, $U=U_A\otimes U_B$, no other quantum evolution can be
accomplished by local means only. Thus almost all quantum
evolutions are non-local.  The main question we address in this
paper is how to describe, qualitatively and quantitatively, the
non-locality of quantum evolutions.

In order to be able to describe the amount of non-locality
contained by the unitary operator $U$ we suggest the following
approach.  We consider that Alice and Bob, in addition of being
able to perform any local operations, they also have {\it
additional resources}, namely they share entangled states, and
they are able to communicate classically. The question then
reduces to finding out how much of these resources is needed to
implement $U$.

The above framework has also been put-forward by Chefles, Gilson
and Barnett \cite{chefles}.

We emphasise that although we have largely discussed the role of
quantum entanglement above, the role of the classical
communication is equally important.  Understanding the character
of a quantum evolution requires knowing both the amount of
entanglement and the amount of classical communication needed.

\section{General sufficiency conditions}

First of all, it is important to note that {\it any} unitary
evolution can be implemented given enough shared entanglement and
classical communication. Indeed, consider the case of two qubits,
one held by Alice and one by Bob. Any unitary transformation $U$
on these two qubits can be accomplished by having Alice teleport
her qubit to Bob, Bob performing $U$ locally and finally Bob
teleporting Alice's qubit back to Alice. The resources needed for
the two teleportation actions are: (1 e-bit plus two classical
bits transmitted from Alice to Bob for the Alice to Bob
teleportation) plus  (1 e-bit plus two classical bits transmitted
from Bob to Alice for the Bob to Alice teleportation). It is
obvious now that any unitary operation involving any number of
parties and any number of qubits can be accomplished by a similar
procedure (teleporting all states to a single location, performing
$U$ locally and teleporting back the qubits to their original
locations).

The ``double teleportation" procedure shown above is {\it
sufficient} to implement any quantum evolution. The question is
however whether so much resources are actually {\it needed}. We
will discuss a couple of specific example below.

\section{The SWAP operation on two qubits}
The SWAP operation defined by:
\begin{eqnarray}
U_{\rm SWAP} | \psi \rangle \otimes | \phi \rangle = |  \phi
\rangle \otimes | \psi \rangle
\end{eqnarray}
is a particularly intriguing case, since although it takes product
states to product states, it is, as we now show, the most
non-local operation possible in the sense described above. That
is, we will prove that in order to implement a SWAP on two qubits
it is not only sufficient but also {\it necessary} to use 2 e-bits
plus 2 bits of classical communication from Alice to Bob plus 2
bits of classical communication from Bob to Alice.

{\it Proof:} To prove that the SWAP operation {\it needs} as
non-local resources 2 e-bits, we will show that if we have an
apparatus able to implement the SWAP operation we can use it in
order to create 2 e-bits. Thus, since entanglement cannot be
created {\it ex nihilo}, the apparatus which implements the SWAP
{\it must} use 2 e-bits as an internal non-local resource.

Let us show how to generate two singlets using the SWAP operation.
Firstly Alice and Bob prepare singlets locally
\begin{eqnarray}
\up_A\up_a + \down_A\down_a \quad \hbox{\rm and} \quad \up_B\up_b
+ \down_B\down_b,
\end{eqnarray}
Alice's spins are labelled $A$ and $a$ and Bob's $B$ and $b$ (here
and in what follows we will leave out normalisation factors for
states). Now perform the SWAP operation on spins $A$ and $B$: $$
(\up_A\up_a + \down_A\down_a)\ ( \up_B\up_b + \down_B\down_b)
\mapsto $$
\begin{eqnarray}(\up_B\up_a + \down_B\down_a)\ ( \up_A\up_b +
\down_A\down_b).
\end{eqnarray}
This state contains two singlets held between Alice and Bob.

To find the classical communication resources {\it needed} to
implement the SWAP operation we will adapt an argument first given
in \cite{teleportation}. We show that if we have an apparatus able
to implement the SWAP operation we can use it in order to
communicate 2 bits from Alice to Bob plus 2 bits from Bob to
Alice. From this follows that it must be the case that the SWAP
apparatus uses 2 bits of classical communication from Alice to Bob
plus 2 bits of classical communication from Bob to Alice as an
internal resource, otherwise Alice could receive information from
Bob transmitted faster than light.

For suppose that the SWAP operation requires less than four bits
of classical communication (two bits each way).  Alice and Bob can
produce an instantaneous SWAP operation which works correctly with
probability greater than one sixteenth in the following way. Alice
and Bob run the usual SWAP protocol, but instead of waiting for
classical communication from each other, they simply guess the
bits that they would have received.  Since we have assumed that
the SWAP operation requires less than 4 bits, the probability that
Alice and Bob guess correctly is greater than one sixteenth and
hence the SWAP operation also succeeds with probability greater
than one sixteenth.

Thus using the protocol described previously can now use this
imperfect, but instantaneous SWAP to communicate 4 bits
instantaneously.  The bits arrive correctly when the SWAP is
implemented correctly. Hence the probability that 4 bits arrive
correctly is larger than one sixteenth; 4 bits communicated
correctly with probability greater than one sixteenth represents a
non-zero amount of information. Thus Alice and Bob have managed to
convey some information to each other instantaneously. We conclude
therefore that the SWAP operation cannot be done with less that 4
bits of classical communication; otherwise it allows communication
faster than the speed of light.

Earlier in this section we showed that the SWAP operation can be
used to generate two singlets.  We now show that the SWAP
operation can be also be used to perform four bits of classical
communication (two bits each way): the main idea is that of
``super-dense coding" \cite{superdense}. Suppose that initially
Alice and Bob share two singlets:
\begin{eqnarray}
\up_A\up_B + \down_A\down_B \quad \hbox{\rm and} \quad \up_a\up_b
+ \down_a\down_b.
\end{eqnarray}
Now Alice chooses one of four local unitary operations 1
(identity), $\sigma_x$, $\sigma_y$,   $\sigma_z$ and performs it
on her spin $A$. This causes the first singlet to be in one of the
four Bell states. Bob also, independently chooses one of these
four locally unitaries and performs it on his spin $b$, putting
the second singlet into one of the Bell states. Then the SWAP
operation is performed on spins $A$ and $b$. Now both Bob and
Alice have one of the Bell states locally; which one they have
depends on which operation the other performed.  By measurement,
they can work out which of the four unitaries the other performed.
Thus the SWAP operation has enabled two bits of classical
communication to be performed each way.

\section{The CNOT operation on two qubits}

Another important quantum operation is CNOT, defined as

\begin{equation}\up\up\mapsto\up\up\end{equation}

\begin{equation}\up\down\mapsto\up\down\end{equation}

\begin{equation}\down\up\mapsto\down\down\end{equation}

\begin{equation}\down\down\mapsto\down\up.\end{equation}

As we prove below, the necessary and sufficient resources for CNOT
are 1 e-bit plus 1 bit of classical communication from Alice to
Bob plus 1 bit of classical communication from Bob to Alice.

{\bf Proof:} {\bf Constructing a CNOT} We now show how to
construct the CNOT operation using one singlet and two bits of
classical communication.  We then show how to generate one singlet
or perform two bits of classical communication using the CNOT.

Firstly we will show how, using one singlet and one bit of
classical communication each way, we can perform a CNOT on  the
state
\begin{eqnarray}
(\alpha\up_A + \beta\down_A)\  (\gamma\up_B +
\delta\down_B)\label{initialproduct}
\end{eqnarray}
i.e. transform it to
\begin{eqnarray}
\alpha\up_A (\gamma\up_B + \delta\down_B) + \beta\down_A
(\gamma\down_B + \delta\up_B).
\end{eqnarray}
Since the operation behaves linearly, the protocol performs the
CNOT on any input state (i.e. even if the qubits are entangled
with each other or with other systems).

{\bf Step 1} The first step is to append a singlet held between
Alice and Bob to the state (\ref{initialproduct}):
\begin{eqnarray}
(\alpha\up_A + \beta\down_A)\ (\up_a\up_b + \down_a\down_b)\
(\gamma\up_B + \delta\down_B),
\end{eqnarray}
then for Alice to measure the total spin of her spins $A$ and $a$.

If the total spin is one, then the state becomes
\begin{eqnarray}
(\alpha\up_A\up_a\up_b + \beta\down_A\down_a\down_b)\ (\gamma\up_B
+ \delta\down_B).\label{spin=one}
\end{eqnarray}

Now Alice disentangles the singlet spin by performing the
following (local) operation:
\begin{eqnarray}
\up_A\up_a \mapsto  \up_A\up_a; \qquad
 \down_A\down_a \mapsto \down_A\up_a,
\end{eqnarray}
and the state becomes
\begin{eqnarray}
(\alpha\up_A\up_b + \beta\down_A\down_b)\ (\gamma\up_B +
\delta\down_B)\up_a.
\end{eqnarray}

If the total spin had been zero, then rather than (\ref{spin=one})
the state becomes
\begin{eqnarray}
(\alpha\up_A\down_a\down_b + \beta\down_A\up_a\up_b)\ (\gamma\up_B
+ \delta\down_B).\label{spin=zero}
\end{eqnarray}
In this case Alice can disentangle  the $a$ spin  by
\begin{eqnarray}
\up_A\down_a \mapsto  \up_A\up_a; \qquad
 \down_A\up_a \mapsto \down_A\up_a,
\end{eqnarray}
leading to
\begin{eqnarray}
(\alpha\up_A\down_b + \beta\down_A\up_b)\ (\gamma\up_B +
\delta\down_B)\up_a.
\end{eqnarray}
In order to get this state in the correct form, Bob needs to
invert his $b$ spin.  Thus Alice must communicate one bit to Bob
to tell him whether she found total spin one or zero, and thus
whether he needs to invert his spin or not.

After these operations, the state is
\begin{eqnarray}
(\alpha\up_A\up_b + \beta\down_A\down_b)\ (\gamma\up_B +
\delta\down_B)\up_a.
\end{eqnarray}

{\bf Step 2} Now Bob performs a CNOT on the $b$ and $B$ spins,
thus the total state is
\begin{eqnarray}
[\alpha\up_A\up_b (\gamma\up_B + \delta\down_B) +
\beta\down_A\down_b (\gamma\down_B + \delta\up_B)]\up_a.
\end{eqnarray}

{\bf Step 3} Bob now measures $\sigma_x$ on his part of the
singlet $b$. Either the state becomes
\begin{eqnarray}
& &[\alpha\up_A (\gamma\up_B + \delta\down_B) + \beta\down_A
(\gamma\down_B + \delta\up_B)]\nonumber\\ & &\qquad \otimes \up_a\
(\up_b + \down_b),
\end{eqnarray}
or
\begin{eqnarray}
& &[\alpha\up_A (\gamma\up_B + \delta\down_B) - \beta\down_A
(\gamma\down_B + \delta\up_B)]\nonumber\\ & &\qquad \otimes\
\up_a(\up_b - \down_b),
\end{eqnarray}

In the former case (i.e. the $x$ component of spin was $+$) we
have performed the protocol as desired.  In the latter, Alice
needs to perform a $\sigma_z$ rotation by $\pi$.  Thus Bob needs
to communicate one bit to Alice to tell her whether or not to
perform the rotation.

We have thus shown how to perform a CNOT using one singlet and one
bit of classical communication each way.

{\bf Creating entanglement by CNOT} We show now that a CNOT
apparatus can be used to create 1 e-bit between Alice and Bob;
thus (since entanglement cannot be increased by local operations)
1 e-bit is a necessary resource for constructing a CNOT.

Creating 1 e-bit by a CNOT is straightforward:
\begin{equation}
(\up_A+\down_A)\up_B\mapsto \up_A\up_B+\down_A\down_B.
\end{equation}

 {\bf Classical communication by CNOT}

Suppose that Alice and Bob have an apparatus which implements a
CNOT and also share 1 e-bit. They can use these resources to
communicate {\it at the same time} 1 classical bit from Alice to
Bob and 1 classical bit from Bob to Alice. This proves (see
preceding section) that communicating 1 classical bit each way is
a necessary resource for constructing a CNOT.

Suppose the initial state is

\begin{equation}\up_a\up_b+\down_a\down_b.\end{equation}

Alice can encode a ``0" by not doing anything to the state and a
``1" by flipping her qubit. Bob can encode a ``0" by not doing
anything to the state and a ``1" by changing the phase as follows:
$\up\rightarrow \up$ and $\down\rightarrow -\down$.

The four states corresponding to the different bit combinations
are thus

 \begin{equation}\up_a\up_b+\down_a\down_b~~corresponds~to~~0_A0_B.\end{equation}

 \begin{equation}\down_a\up_b+\up_a\down_b~~corresponds~to~~1_A0_B.\end{equation}

 \begin{equation}\up_a\up_b-\down_a\down_b~~corresponds~to~~0_A1_B.\end{equation}

 \begin{equation}\down_a\up_b-\up_a\down_b~~corresponds~to~~1_A1_B.\end{equation}

After encoding their bits, Alice and Bob apply the CNOT operation.
This results in the corresponding four states

 \begin{equation}\up_a\up_b+\down_a\up_b=(\up_a+\down_a)\up_b~~corresponds~to~~0_A0_B\end{equation}

 \begin{equation}\down_a\down_b+\up_a\down_b=(\up_a+\down_a)\down_b~~corresponds~to~~1_A0_B\end{equation}

 \begin{equation}\up_a\up_b-\down_a\up_b=(\up_a-\down_a)\up_b~~corresponds~to~~0_A1_B\end{equation}

 \begin{equation}\down_a\down_b-\up_a\down_b=(\down_a-\up_a)\down_b~~corresponds~to~~1_A1_B.\end{equation}
Bob can now find out Alice's bit by measuring his qubit in the
\{$\up_b$, $\down_b$\} basis while Alice can find out Bob's bit by
measuring her qubit in the \{$\up_a+\down_a$, $\up_a-\down_a$\}
basis.

\section{The Double CNOT operation on two qubits}

    One might have thought that the SWAP operation was the {\it unique} maximally
non-local operation, at least in the terms used in this paper.  We
here demonstrate that there is another maximally non-local
operator, which is the ``Double CNOT", or ``DCNOT" gate, formed by
acting a CNOT from particle 1 onto particle 2, and then a second
CNOT from particle 2 onto particle 1. It is defined by

\begin{equation}\up\up\mapsto\up\up\end{equation}

\begin{equation}\up\down\mapsto\down\down\end{equation}

\begin{equation}\down\up\mapsto\up\down\end{equation}

\begin{equation}\down\down\mapsto\down\up.\end{equation}

To show that DCNOT is maximally non-local, we shall first
demonstrate that it can be used to create 2 e-bits. We shall then
show that it can be used to communicate 2 bits of information from
Alice to Bob, and simultaneously to send 2 bits from Bob to Alice.
The argument used for the SWAP operation then proves that to build
a DCNOT we need 2 e-bits plus 2 bits of classical communication
from Alice to Bob plus 2 bits of classical communication from Bob
to Alice. Since any transformation on two qubits can be performed
using these resources via teleportation, we will then have shown
that the DCNOT is maximally non-local, in terms of resources.

Creating 2 e-bits is easy.  Alice and Bob prepare singlets
locally, and then perform the DCNOT on spins $A$ and $B$:

$$ (\up_A\up_a + \down_A\down_a)\ ( \up_B\up_b + \down_B\down_b) \mapsto $$
\begin{eqnarray} \up_A\up_a\up_B\up_b + \down_A\up_a\down_B\down_b +
\up_A\down_a\down_B\up_b + \down_A\down_a\up_B\down_b.
\end{eqnarray}

We now have a Schmidt decomposition of rank 4, ie. a 2 party state which is
locally equivalent to 2 e-bits between Alice and Bob.

Transmitting 2 bits of information in both directions at the same
time is a little more tricky.  Alice and Bob need to have 2 e-bits
in addition to the DCNOT operation.  They first transform  their
e-bits (locally) into the state

\begin{equation} \up_A\up_a\up_B\up_b + \down_A\up_a\up_B\down_b + \down_A\down_a\down_B\up_b + \up_A\down_a\down_B\down_b.  \end{equation}

Alice now encodes 1 bit of information in the state by either
applying, or not applying $\sigma_z\otimes\sigma_z$ to her 2
spins.  She encodes a second bit of information by applying, or
not applying $\sigma_x$ to her first spin, $A$.  Bob similarly
encodes two bits of information, using the transformation
$\sigma_z$ on spin $B$ to encode his first bit, and
$\sigma_x\otimes\sigma_x$ to encode his second bit.

Having encoded the information, they make it locally accessible by
applying the DCNOT to spins $A$ and $B$.  It is not obvious, but
simple to check, that Alice and Bob now each have one of the 4
Bell states locally, and that Alice's particular state corresponds
to Bob's encoded bits, and vice-versa.

\section{Multi-partite operations}

In the previous sections we studied different bi-partite
operations. What about multi-partite operations, such as the
Toffoli or the Fredkin gates on three qubits? As we showed in
section II, they can all be implemented by using the ``double
teleportation" method. On the other hand, finding the {\em
necessary} resources is far more difficult than in the bi-partite
case; indeed it is not possible at present. The reason is that
there exist different inequivalent types of multi-partite
entanglement \cite{multipartite,schumacher}. For example, it is
known that singlets and GHZ states are inequivalent in the sense
that they cannot be reversibly transformed into each other, not
even in the asymptotic limit. Although GHZs (as all other
entangled states) can be built out of singlets, such a procedure
is wasteful. Hence, when investigating the minimal entanglement
resources needed to implement multi-partite quantum operations, we
have to use the different inequivalent types of entanglement.
Unfortunately, at present multi-partite entanglement is far from
being fully understood.

\section{``Conservation" relations}

In studying the non-locality of quantum states a most important
issue is that of ``manipulating" entanglement, i.e. of
transforming some states into others \cite{concentration}.
Similarly we can ask: Given a unitary evolution, can we use it to
implement some other unitary evolution?

In particular, for pure quantum states we have {\it conservation}
relations \cite{concentration,thermodynamics}. For example, when
Alice and Bob share a large number $n$ of pairs of particles, each
pair in the same state $\Psi$, they could use these pairs to
generate some other number, $k$, of pairs in some other state
$\Phi$. In the limit of large $n$, this transformation can be
performed {\it reversibly}, meaning that the total amount of
non-locality contained in the $n$ copies of the state $\Psi$ is
the same as the total amount of non-locality contained in the $k$
copies of the state $\Phi$. Is something similar taking place for
unitary transformations?

For unitary transformations we have not yet studied the case of
the asymptotic limit, i.e. performing the same transformation $U$
on many pairs of particles. However, an interesting pattern
emerges even at the level of a single copy.

Consider first the case of SWAP. We know what  the minimal
resources needed to implement a SWAP are. But suppose now that we
are given a device which implements a SWAP. Could we could use it
to get back the original resources needed to create the SWAP?

The balance of resources needed to implement a SWAP can be written
as

\begin{equation}2 \hbox{e-bits} + 2 \hbox{bits}_{A\rightarrow B} + 2 \hbox{bits}_{B\rightarrow A}
=> \hbox{SWAP}.\end{equation} The question is whether

\begin{equation}\hbox{SWAP}=>2 \hbox{e-bits} + 2 \hbox{bits}_{A\rightarrow B} + 2 \hbox{bits}_{B\rightarrow A}
?\end{equation}

Though we do not have yet a complete proof, it appears that the
answer to the above question is ``No". That is, combining
entanglement and classical communication resources to yield a SWAP
is an {\it irreversible process} - we cannot use the SWAP to get
the resources back.

On the other hand, looking back to the proof of the resources
needed for SWAP, we see that we can write the following tight
``implications":

\begin{equation}2 \hbox{e-bits} + 2 \hbox{bits}_{A\rightarrow B} + 2 \hbox{bits}_{B\rightarrow A} => 1
\hbox{SWAP}.\end{equation}

\begin{equation}2 \hbox{e-bits} + 1 \hbox{SWAP} =>  2 \hbox{bits}_{A\rightarrow B} + 2
\hbox{bits}_{B\rightarrow A}.\end{equation}

\begin{equation}1 \hbox{SWAP} =>2 \hbox{e-bits}.\end{equation}

The first of these three implications is to be read as ``given
$2\hbox{e-bits}$ and $ 2 \hbox{bits}_{A\rightarrow B}$ and $ 2
\hbox{bits}_{A\rightarrow B}$ we can produce the SWAP operation;
also if we wish to produce the SWAP operation with e-bits, and
bits communicated from Alice to Bob and vice-versa, we cannot do
so with fewer than $2\hbox{e-bits}$ and $ 2
\hbox{bits}_{A\rightarrow B}$ and $ 2 \hbox{bits}_{A\rightarrow
B}$."

The second and third implications have a slightly different
meaning.  For example we read the second implication as ``given 1
SWAP and 2 e-bits, we can communicate 4 classical bits (two each
way); also we cannot communicate more than 4 classical bits (two
each way) ". On the other hand, it does not mean that ``1 SWAP and
2 e-bits are necessary for communicating 4 classical bits (two
each way) " - for example we can implement this classical
communication with 2 SWAPs.

Exactly the same implications apply for the DCNOT.

\begin{equation}2 \hbox{e-bits} + 2 \hbox{bits}_{A\rightarrow B} + 2 \hbox{bits}_{B\rightarrow A} => 1
\hbox{DCNOT}.\end{equation}

\begin{equation}2 \hbox{e-bits} + 1 \hbox{DCNOT} =>  2 \hbox{bits}_{A\rightarrow B} + 2
\hbox{bits}_{B\rightarrow A}.\end{equation}

\begin{equation}1 \hbox{DCNOT} =>2 \hbox{e-bits}.\end{equation}

Furthermore, very similar implications can be written for the
CNOT:

\begin{equation}1 \hbox{e-bit} + 1 \hbox{bit}_{A\rightarrow B} + 1 \hbox{bit}_{B\rightarrow A} => 1
\hbox{CNOT}.\end{equation}

\begin{equation}1 \hbox{e-bit} + 1 \hbox{CNOT} =>  1 \hbox{bit}_{A\rightarrow B} + 1 \hbox{bit}_{B\rightarrow
A}.\label{ebit+cnot}\end{equation}

\begin{equation}1 \hbox{CNOT} =>1 \hbox{e-bit}.\label{cnot-ebit}\end{equation}

In fact these implications are very similar to the implications
which describe teleportation and super-dense coding which appear,
together with many other similar implications on Bennett's famous
transparency presented at almost all early quantum information
conferences:

\begin{equation}1 \hbox{e-bit} + 2 \hbox{bits}_{A\rightarrow B} => 1 \hbox{qubit}\label{Bennett-relation1}\end{equation}

\begin{equation}1 \hbox{e-bit} + 1 \hbox{qubit}  => 2 \hbox{bits}_{A\rightarrow B} \label{Bennett-relation2}\end{equation}

\begin{equation}1 \hbox{qubit}  => 1 \hbox{e-bit} \label{Bennett-relation3}\end{equation}

The above three implications
(\ref{Bennett-relation1},\ref{Bennett-relation2},\ref{Bennett-relation3})
are generally thought to describe relations between classical
information, quantum information and entanglement. However, we
would like to argue that their true meaning is may be more closely
related to dynamics, and that a more illuminating form is probably

\begin{equation}1 \hbox{e-bit} + 2 \hbox{bits}_{A\rightarrow B} =>1 \hbox{teleportation}_{A\rightarrow
B}\end{equation}

\begin{equation}1 \hbox{e-bit} +1 \hbox{teleportation}_{A\rightarrow B}  => 2 \hbox{bits}_{A\rightarrow
B} \end{equation}

\begin{equation}1 \hbox{teleportation}_{A\rightarrow B}  => 1 \hbox{e-bit} \end{equation}

 We conjecture that similar relations hold
between any quantum action and the resources needed to implement
it, that is

\begin{equation}Entanglement + Classical Communication => Action \end{equation}

\begin{equation}Entanglement + Action   => Classical Communication \end{equation}

\begin{equation}Action  => Entanglement \end{equation}

It may be that these relations hold, in general, only in the
asymptotic limit of many copies of the quantum action.

\section{Different ways of achieving the same task}

It is interesting to note that although the transformation from
resources to unitary actions is irreversible, sometimes the same
end product can be achieved in two different ways. For example,
there are two alternative ways to implement

\begin{equation} 2 \hbox{CNOTs} => 1 \hbox{bit}_{A\rightarrow B} + 1 \hbox{bit}_{B\rightarrow
A}.\end{equation}

The first way is to use one CNOT to transmit 1 classical bit from
Alice to Bob and the other CNOT to transmit  1 classical bit from
Bob to Alice, i.e.

\begin{equation} 1 \hbox{CNOT} => 1 \hbox{bit}_{A\rightarrow B}\end{equation} and

\begin{equation} 1 \hbox{CNOT} => 1 \hbox{bit}_{B\rightarrow A}.\end{equation}

Another possibility is to use first one CNOT to create 1 e-bit
(\ref{cnot-ebit}) then the other CNOT plus the e-bit to transmit
the 2 classical bits (\ref{ebit+cnot}), i.e.

\begin{equation} 2 \hbox{CNOTs} => 1 \hbox{e-bit} +1 \hbox{CNOT} => 1 \hbox{bit}_{A\rightarrow B} + 1
\hbox{bit}_{B\rightarrow A}.\end{equation}

\section{Catalysing classical communication}

A very interesting phenomenon is that of ``catalysing" classical
communication. This phenomenon is similar in its spirit to that of
``catalysing entanglement manipulation"
\cite{jonathan,multipartite}. An example is the following.

On its own, the SWAP can only send one bit in each direction at
the same time, and cannot be used for Alice to send 2 bits to Bob,
even if Bob sends no information whatsoever. That is,

\begin{equation}1 \hbox{SWAP} \neq>  2 \hbox{bits}_{A\rightarrow B}.\end{equation}

However, if Alice and Bob share 1 e-bit, Alice can send 2 bits to
Bob {\it without destroying} the e-bit, i.e.
\begin{equation}1 \hbox{SWAP}+1\hbox{e-bit} =>  2 \hbox{bits}_{A\rightarrow B}
+1\hbox{e-bit}.\label{catalysed-communication}\end{equation}

This may be done as follows.  Initially Alice and Bob share a
non-local singlet; Bob also prepares a second singlet locally.
Alice encodes the two bits she wishes to send to Bob by performing
one of the four rotations 1, $\sigma_x$, $\sigma_y$, $\sigma_z$ on
her half of the non-local singlet.  By performing the SWAP
operation on Alice's particle from the non-local singlet and one
particle of the singlet that Bob has prepared locally, Alice and
Bob end up with a non-local singlet held between them; also Bob
can find out the two bits by measurements on the local singlet he
now holds. Specifically, we begin with the state:

\begin{equation}
(\up_A \up_{b1} + \down_A \down_{b1})(\up_B \up_{b2} + \down_B
\down_{b2}),
\end{equation}
where $A$ is Alice's particle, and $B$, $b1$ and $b2$ are Bob's
particles.  Alice performs one of the rotations 1, $\sigma_x$,
$\sigma_y$, $\sigma_z$ on her particle.  They then perform the
SWAP on particles $A$ and $B$, and get (if Alice performed 1):

\begin{equation}
(\up_B \up_{b1} + \down_B \down_{b1})(\up_A \up_{b2} + \down_A
\down_{b2})
\end{equation}

If Alice performed one of the other rotations, Bob will get one of
the other Bell states in system ($B$, $b1$).  Bob now measures
that system in the Bell basis to extract the information, and
Alice and Bob are left with a singlet between systems $A$ and
$b2$.

In effect the SWAP acts as a double teleportation; one from Alice
to Bob and one from Bob to Alice.  Teleporting Alice's qubit, in
conjunction with the e-bit, implements a transmission of two bits
from Alice to Bob using super-dense coding; it destroys the e-bit
in the process.  Simultaneously, the Bob to Alice teleportation
restores the e-bit.

\section{Trading one type of actions for another}

An interesting question is the following. There are cases in which
two different actions require the same resources. For example the
resources needed for 1 SWAP are the same as for 2 CNOTs, i.e., $2
\hbox{e-bits} + 2 \hbox{bits}_{A\rightarrow B} + 2
\hbox{bits}_{B\rightarrow A}$. Now, suppose we had already used
the resources to build 2 CNOTs, but we wanted to change our mind
and we wanted to do 1 SWAP instead. Due to the irreversibility
discussed above, we cannot simply get back the original resources
and use them to construct the SWAP. Is it however possible to go
{\it directly} from 2 CNOTs to 1 SWAP, without going back to the
original resources?  As far as we are aware, the answer is ``No".

It turns out however that if we have many CNOTs  it is
nevertheless useful to build a SWAP from CNOTs directly rather
than going back to the original resources. Indeed, to obtain the
entanglement and classical communication resources needed for 1
SWAP, i.e. $2 \hbox{e-bits} + 2 \hbox{bits}_{A\rightarrow B} + 2
\hbox{bits}_{B\rightarrow A}$ we need 4 CNOTs. However, it is
well-known that  one can construct  1 SWAP directly from 3 CNOTs.
Indeed, we don't even need 3 CNOTs, but can realize a SWAP by

\begin{equation}2 \hbox{CNOTs} +  1 \hbox{bit}_{A\rightarrow B} + 1 \hbox{bit}_{B\rightarrow A} =>1
\hbox{SWAP}\end{equation} which uses less non-local resources than
3 CNOTs. To see this, it suffices to note that

\begin{equation}1 \hbox{CNOT} + 1 \hbox{bit}_{A\rightarrow B} =>1 \hbox{teleportation}_{A\rightarrow
B}\label{teleportationAB}\end{equation} and similarly

\begin{equation}1 \hbox{CNOT} + 1 \hbox{bit}_{B\rightarrow A} =>1 \hbox{teleportation}_{B\rightarrow
A}\end{equation}

To implement (\ref{teleportationAB}) Alice starts with her qubit
in the state $\Psi=\alpha \up+\beta\down$ which has to be
teleported and Bob with his qubit in the state $\up$. After CNOT
the state becomes:

\begin{equation}\Psi\up=(\alpha \up+\beta\down)\up\mapsto \alpha
\up\up+\beta\down\down\end{equation}

Alice then measures her qubit in the $|+>={1\over{\sqrt
2}}(\up+\down)$ and $|->={1\over{\sqrt 2}}(\up-\down)$ basis and
communicates the result to Bob. If $(+)$ then Bob's qubit is
already in the required state $\Psi=\alpha \up+\beta\down$; if
$(-)$ then Bob's qubit is in the state $\Psi'=\alpha
\up-\beta\down$ and Bob can obtain $\Psi$ by changing  the
relative phase between $\up$ and $\down$  by $\pi$.

\bigskip\bigskip
\noindent{\em Note added.} While completing this work we became
aware of closely related work by J. Eisert, K. Jacobs, P.
Papadopoulos and M. Plenio \cite{plenio}.

\end{multicols}
\end{document}